\documentclass[final,3p,times,twocolumn]{elsarticle}
\usepackage{hyperref}
\usepackage{verbatim}
\usepackage{url}
\usepackage{gensymb}
\usepackage{amsmath}

\usepackage{multirow}
\usepackage{multicol}

\usepackage{amssymb}

\usepackage{lineno}[displaymath, mathlines]
\usepackage{subcaption}
\usepackage{siunitx}

\journal{Nuclear Instrument and Methods in Physics}

\begin{document}

\begin{frontmatter}



\title{Radiation damage on SiPMs for Space Applications}


\author[fbk,uniud,tifpa]{Anna Rita Altamura} 
\author[fbk,tifpa]{Fabio Acerbi} 
\author[tifpa]{Benedetto Di Ruzza}
\author[tifpa]{Enrico Verroi}
\author[fbk]{Stefano Merzi}
\author[fbk,tifpa]{Alberto Mazzi}
\author[fbk,tifpa]{Alberto Gola}

\address[fbk]{Fondazione Bruno Kessler (FBK), Sensors \& Devices (SD), via Sommarive 18, I-38123, Trento, Italy}
\address[uniud]{University of Udine, via Palladio 8, 33100 Udine, Italy}
\address[tifpa]{TIFPA - Trento Institute for Fundamental Physics and Applications, via Sommarive 14, I-38123, Trento, Italy}
\address[unitn]{University of Trento, via Sommarive 14, I-38123, Trento, Italy}

\begin{abstract}
Silicon Photo-multipliers (SiPMs) are very sensitive photo-detectors that experienced a big development in the last years in several applications, like LIDAR, astrophysics, medical imaging and high energy physics (HEP) experiments. In HEP experiments, in particular, they are often exposed to significant radiation doses. 
The main purpose of this manuscript is the characterization of several FBK SiPM technologies when exposed to 74 $MeV$ protons with a total fluence comparable to the one that they would experience in space along circular Low Earth Orbits (LEO), Polar, during a five years mission. 

In this work, we estimated the expected proton fluences along the selected orbit, by means of the SPENVIS software. Several fluence steps were chosen to consider dense fluence intervals and have a more accurate sight on the whole damage process. We estimated a maximum fluence achieved during the tests of $6.4 \times 10^{11}$ $n_{eq}/cm^2$.
Based on such simulations, we irradiated several SiPM technologies. We developed a custom experimental setup, which was used to perform online reverse voltage-current, right after each irradiation step, to minimize the effect of the annealing on the measurement. 

The results are then displayed, in particular the currents, the noise and the Photon Detection Efficiency. Also a 30-days study on the annealing of the devices was performed. 

Lastly, the conclusions are drawn on the basis of the Signal-to-Noise Ratio (SNR), taking into account the standard parameters of famous satellites using similar orbits as the ones considered into this work.  

\end{abstract}


\end{frontmatter}


\section{\textit{Introduction}}
Silicon Photo-multipliers (SiPMs) are solid-state detectors working in Geiger mode based on an array of single-photons avalanche diodes (SPADs), thus with an internal gain which enables them to have a good single-photon resolution. They have experienced a large performance improvement over the last few years and they are used in several fields, from nuclear medicine\cite{nuclear}, high energy physics\cite{CMSHCAL}, automotive, LIDAR\cite{automotive}. Space is one of the applications where they have been less explored so far\cite{space}, although several experiments are starting to employ them\cite{Space_sipm}. One of the crucial limits of using SiPMs in HEP and space applications is their radiation tolerance. Indeed, when exposed to a huge amount of radiation, SiPMs could suffer a significant damage which, depending on the type of radiation, might happen either into the bulk or on the surface of the device\cite{Moll}. 
In this sense, this work aims to evaluate the performance of several FBK technologies after being exposed to proton fluences comparable with the ones that are expected in space environment for satellites along some of the most used orbits in space applications. 
In particular, we irradiated the SiPMs at the Proton Therapy Center in Trento with 74 $MeV$ protons, at fluences between $7.4\times 10^6$ $n_{eq}/cm^2$ and  $6.4\times10^{11}$ $n_{eq}/cm^2$.

Several FBK technologies were tested: VUV-HD 2019\cite{vuv}, NUV-HD-RH, NUV-HD-cryo\cite{nuv}, RGB-HD and RGB-HD "EnhancedBorder" \cite{rgb}. The tested structures included several $1\times1$ $mm^2$ SiPMs manufactured with the same technology, different cell pitches and structure. The different structures were assembled on a PCB, which will be described in detail later, and placed along a ring to obtain an irradiation as uniform as possible on each SiPM. 

In the next sections, we will introduce the fluence values estimated from simulations. Subsequently, we will show the experimental setup during the irradiation tests, the results of the characterization measurements with a focus on the comparison among the technologies under test and then the conclusion will be derived in the last section. 

\section{Fluence Range Calculation}
\begin{figure}[tb]
    \centering
    \includegraphics[width=0.45\textwidth]{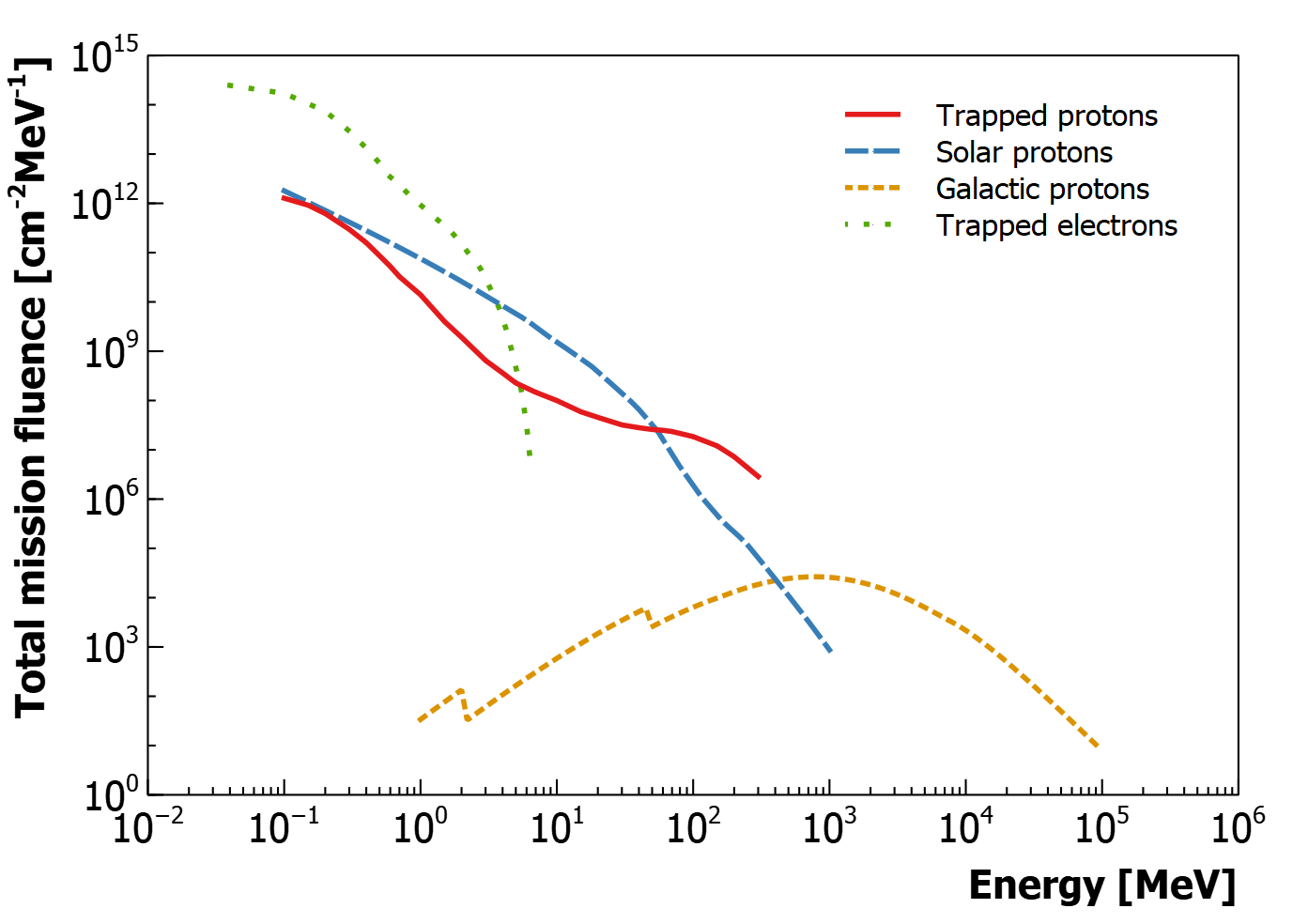}
    \caption{ Output plot from SPENVIS software showing the five years total mission particle fluence as a function of the energy for trapped, solar, galactic protons and trapped electrons, along a ”HERMES-like”" circular Low Earth Orbit, Polar (550 $km$ - 97$^\circ$).}
    \label{fig:fluence}
\end{figure}

\begin{figure}[tb]
    \centering
    \includegraphics[width=0.45\textwidth]{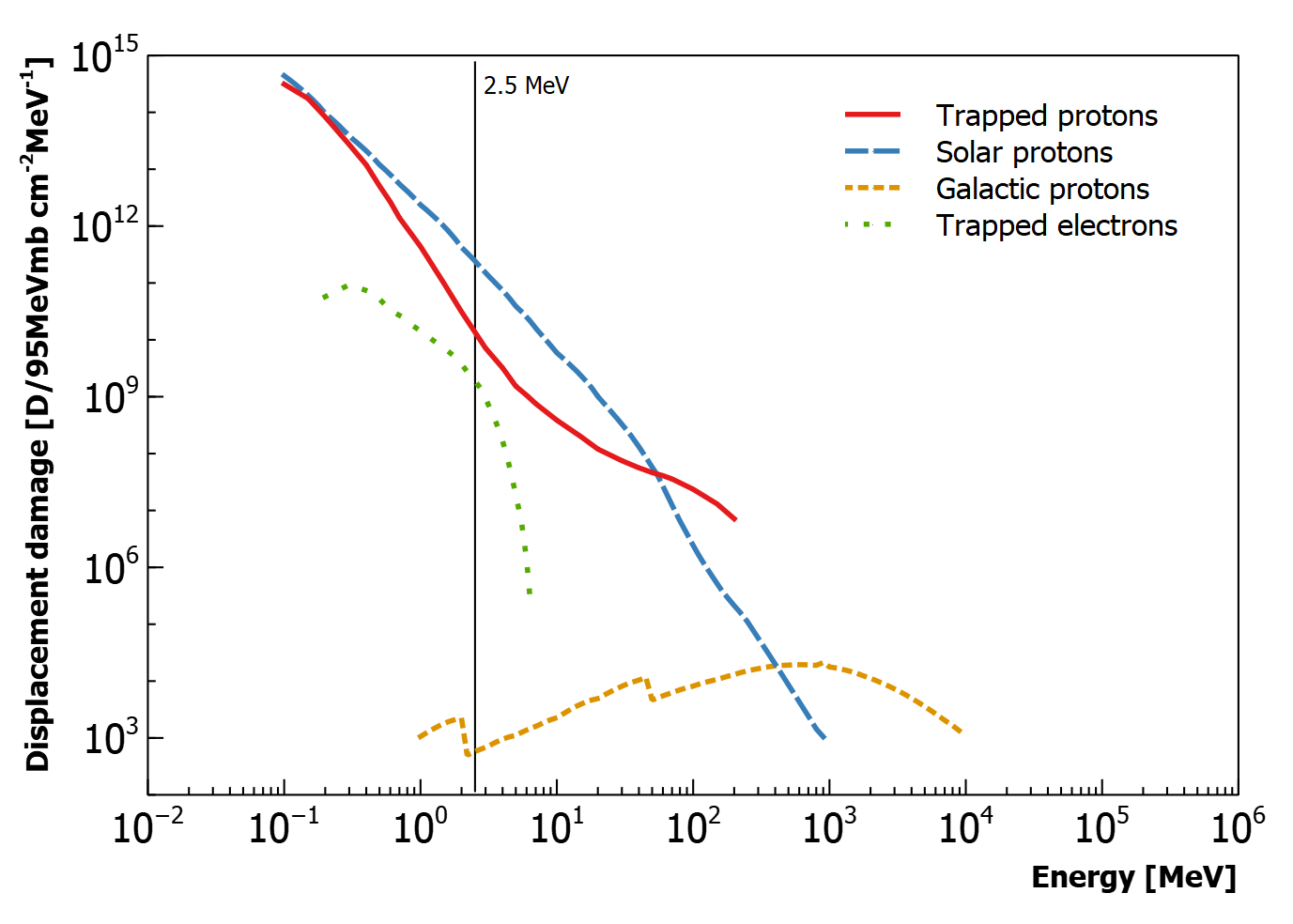}
    \caption{Total damage of trapped, solar and galactic protons and trapped electrons over the whole mission lifetime. The line marked at 2.5 MeV represents the energy cut which is assumed to be applied to data.}
    \label{fig:totaldamage}
\end{figure}

\begin{figure}[tb]
    \centering
    \includegraphics[width=0.45\textwidth]{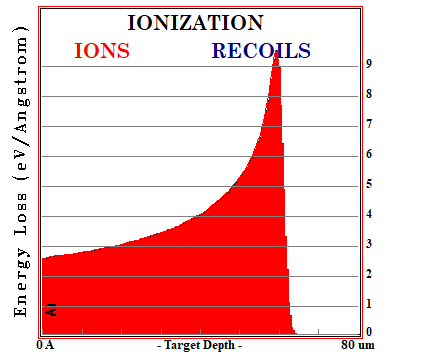}
    \caption{SRIM simulation of 2.5 MeV H ions energy loss in 80 $\mu$m Al target}
    \label{fig:srim}
\end{figure}

In our study, a typical satellite orbit was considered: 550 $km$ - 97$^\circ$ inclination (circular Low Earth Orbits, Polar)\cite{HERMES-SP:2021hvq}. The expected proton fluence along this orbit was estimated through the use of SPENVIS software\cite{spenvis}, for a five years mission starting on July 1st 2022 with maximum solar activity. SPENVIS provided as an output the spectra of total mission average flux $[cm^{-2}s^{-1} MeV^{-1}]$ and total mission fluence $[cm^{-2}MeV^{-1}]$ related to several contributions: protons and electrons trapped into the terrestrial magnetic field, solar protons and galactic protons. 
In Fig.\ref{fig:fluence} the total mission particle fluences as a function of the protons energy for trapped, solar, galactic protons and trapped electrons are visible along the selected circular polar LEO orbit. 
The fluence spectrum of each contribution (even from different particle types) was then converted into an equivalent spectrum of 1 MeV neutron damage in Silicon, enabling us to compare them. To do that, the NIEL scaling hypothesis was considered. For each energy value of the spectrum $E$, the total displacement damage $DD(E)$ corresponding to the proton fluence $\phi(E)$ was calculated as:

\begin{equation}
    DD(E) = D(E) \cdot \phi(E)
\end{equation}

where $D(E)$ is the displacement damage cross section according to the NIEL scaling hypothesis, which converts the effective particle damage into a 1 MeV neutron equivalent damage. A plot of this spectrum is shown in Fig.\ref{fig:totaldamage}, where we can clearly observe that the solar and trapped protons have the highest impact on the total damage in Silicon, whereas the galactic protons and the trapped electrons appear to affect it less. For this reason we chose to neglect these two latter contributions, taking into account only the most relevant two. 
Assuming a 80$\mu$m-thick Aluminum shielding of the whole detector, this produces a cut-off of the protons with energy lower than 2.5 MeV and reduces the total amount of damage on the sensors. 
SRIM software \cite{srim} was used for the Aluminium shielding simulations. The resulting 2.5 MeV H ions energy loss in 80 $\mu$m Al target is visible in Fig.\ref{fig:srim}, where a total absorption of the 2.5 MeV protons is visible at about 65 $\mu$m, thus a 80 $\mu$m Al shielding results to be a reasonable choice to cut all the protons with energy lower than 2.5 MeV.

The resulting equivalent damage spectrum was then integrated over the whole protons energy range $[2.5 MeV, E_{max}]$:

\begin{equation}
    D_{tot} = \int_{2.5 MeV}^{E_{max}} DD(E) dE
\end{equation}

\begin{figure}[tb]
    \centering
    \includegraphics[width=0.45\textwidth]{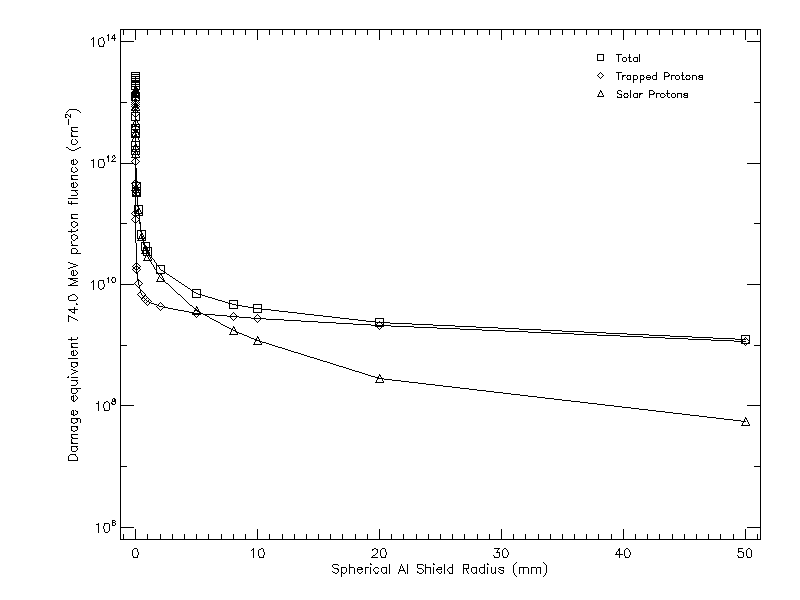}
    \caption{Equivalent 74 $MeV$ protons fluence as a function of the radius of an Aluminum spherical shielding along a Low Polar Orbit (550 $km$ - 97$^\circ$) for trapped and solar protons.}
    \label{fig:damagespectrum}
\end{figure}

The final step was to extract the proton fluence at 74 MeV required to create such a total damage which means to normalize the total damage to the damage of 74 MeV protons, according to the equation: 
\begin{equation}
    \phi (74 MeV)= \frac{\int D(E) \phi(E) dE} {D(74 MeV)} = \frac{D_{tot}}{D(74 MeV)}
\end{equation}

where $D(E)$ is the 1 MeV neutron equivalent damage of the protons and $D(74 MeV)$ is the 1 MeV neutron equivalent damage of 74 MeV protons. In Fig.\ref{fig:damagespectrum}, the 74 MeV equivalent protons fluence from SPENVIS software related to trapped, solar and total protons is plotted as a function of the Al shield radius. Here, we can observe that at 80$\mu$m, we are at $4\times 10^{11}$ 74 MeV total protons, which means $6\times10^{11}$ $n_{eq}/cm^2$.

The fluence range considered in this work is $7.4\times 10^4 \div 6.4\times 10^{11} $ $n_{eq}/cm^2$. 

The final proton irradiation fluences on the sensors were estimated integrating the beam profiles on each SiPM, taking into account its specific position within the beam spot. 

A set of SiPMs was irradiated at $10^{11}$ $n_{eq}/cm^2$. This enabled us to have a set of devices for further measurements apart from the parameters estimated through the online current measurements. These SiPMs were kept in a refrigerate environment ($-10^{\circ}C$), to avoid as much as possible the annealing process.

\section{Experimental setup}
At the Trento Proton Therapy Center there are two possible irradiation systems\cite{DiRuzza:2021oq} with two different beam settings: a dual-ring double scattering system\cite{dualring} which ensures a large field irradiation at 148 MeV with a $98\%$ uniformity on a circular area with an approximately 5.9 cm diameter and a gaussian-profiled beam \cite{gaussian} with a 7 mm $\sigma$ and a large energy range. In particular, we decided to use the dual-ring system which provided an higher uniformity. In this case, the standard 148 MeV energy was lowered to 74 MeV through a Water-equivalent polystyrene material (RW3) resulting in a beam with an equivalent uniform circular area with 7.8 cm radius. 

\begin{figure}[tb]
    \centering
    \includegraphics[width=0.45\textwidth]{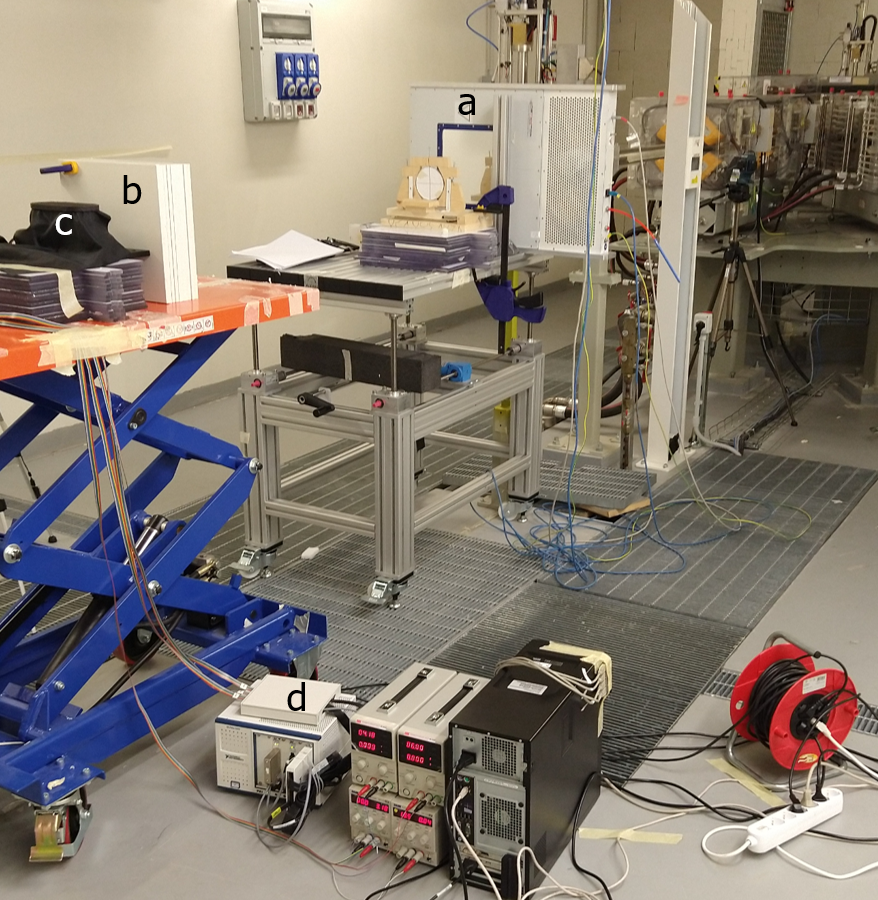}
    \caption{Irradiation setup with the double-ring scattering system at the Proton Therapy Center in Trento, where the different setup components are clearly visible: the ionization chamber (a), the PW3 used for the energy lowering of the beam (b), the dark box including the irradiated sensors (c) and the I-V measurement setup (d).}
    \label{fig:setup}
\end{figure}

The experimental setup included the beam output, an ionization chamber to measure the total number of protons passing through it and a \textit{dark box}, which contained the devices during the irradiation and the measurements. In Fig.\ref{fig:setup} the setup for the dual-ring system is shown, where the dual ring and the RW3 blocks to lower the ions energy are visible. 

\begin{figure}[tb]
    \centering
    \includegraphics[width=0.35\textwidth]{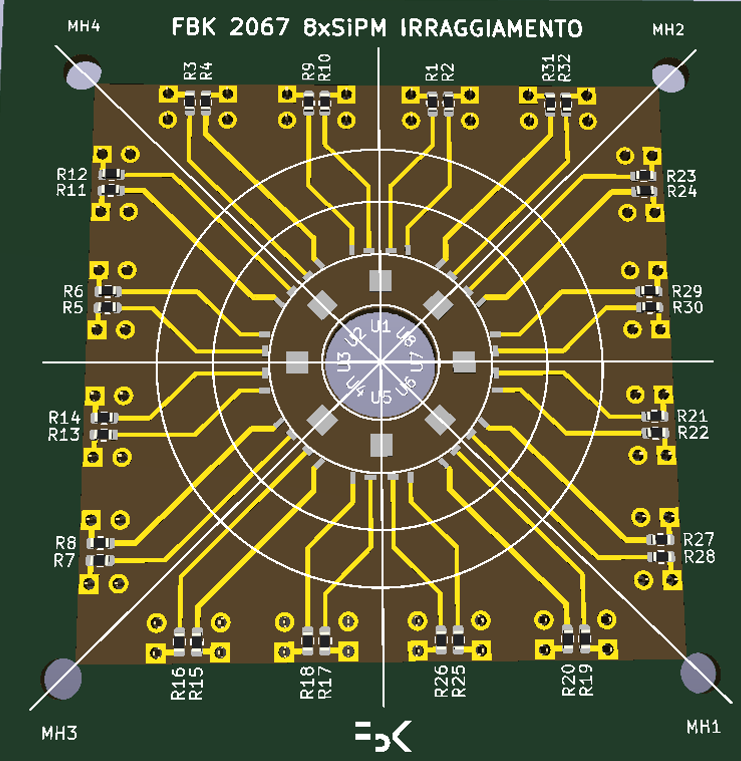}
    \caption{Layout image of the PCB used for the irradiation test}
    \label{fig:PCB}
\end{figure}

The devices were irradiated in several steps and reverse current-voltage (I-V) measurements were performed after each irradiation step to reduce the impact of the annealing and to isolate the results of the damage effects.
The devices were placed on a custom Printed Circuit Board (PCB), where the SiPMs were mounted for the measurement, as shown in Fig.\ref{fig:PCB}. The PCB included eight spots. SiPMs on each structure were connected to the PCB with wire bonding and their outputs connected through 2 m long cables to the measurement system. During the whole irradiation time, two PCB where irradiated, each with a different final fluence: $1.4\times 10^8$ $n_{eq}/mm^2$ and $6.4\times 10^9$ $n_{eq}/mm^2$. PCBs were inserted inside a custom dark box realized by 3D plastic printing, with a motor-controlled light shutter. This was open during the irradiation stage. After each irradiation step, the shutter was closed and all the SiPMs were measured in dark conditions.

After irradiating and measuring the last PCB, which was the one with the highest final fluence, the devices was kept in the same dark box and left to anneal at room temperature ($20\div 25^{\circ}C$) for approximately 30 days. During this time reverse I-V measurements were performed twice a day. 

A set of SiPMs was irradiated as naked chips at a fluence of $10^{11}$ $n_{eq}/cm^2$ and then mounted on a PCB. This enabled us to have a set of SiPMs irradiated at a lower fluence available for measurements of characterization.

\section{Results}

\subsection{Reverse Current and Breakdown voltage}

\begin{figure*}[tb]
	\centering
	\begin{subfigure}{.47\textwidth}
		\includegraphics[width=1\textwidth]{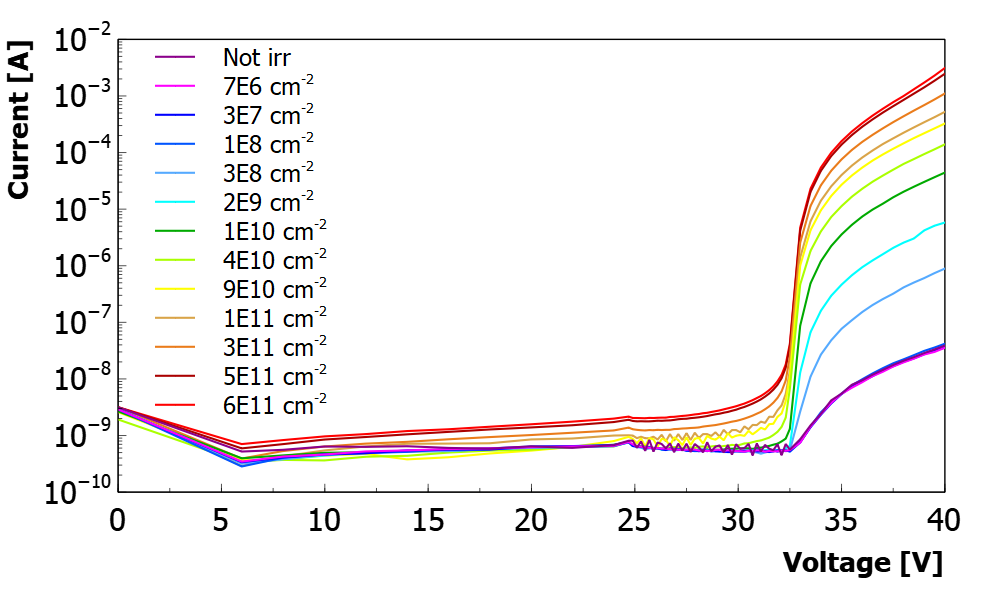}
	\end{subfigure}
\hspace{0.7cm}
	\begin{subfigure}{.47\textwidth}
		\includegraphics[width=1\textwidth]{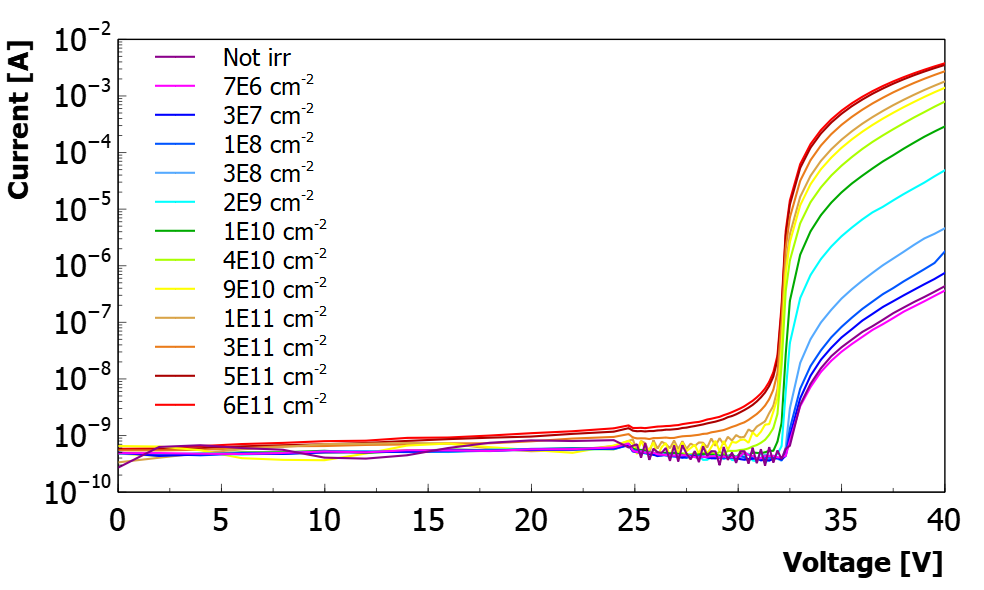}
	\end{subfigure}
	\caption{Reverse current as a function of the bias voltage at the several fluences for the NUV-HD-RH SiPM with 20$\mu$m cell (left) and the VUV-HD 2019 with 35$\mu$m cell (right) in dark conditions.}
	\label{fig:IV}
\end{figure*}

The reverse current is one of the key parameters for the characterization of a SiPM. In fact, when irradiated, the induced defects result in an increase of both the "leakage current" (not multiplied), from the surface and the "dark current" (multiplied), from the bulk of the SiPM \cite{altamura2021characterization}. The increase of these parameters provide an indication of the quantity of damage that the sensor has experienced. In this test, the measurement of the reverse current as a function of the applied voltage was taken shortly after each irradiation step. Each measurement was performed both in dark and light conditions through a current-controlled LED inside the dark box. In Fig.\ref{fig:IV} the increase of the reverse current with fluence can be observed for the NUV-HD-RH SiPM with 20 $\mu$m cell and the VUV-HD 2019 with 35 $\mu$m cell in dark conditions. 

\begin{figure}[tb]
    \centering
    \includegraphics[width=0.45\textwidth]{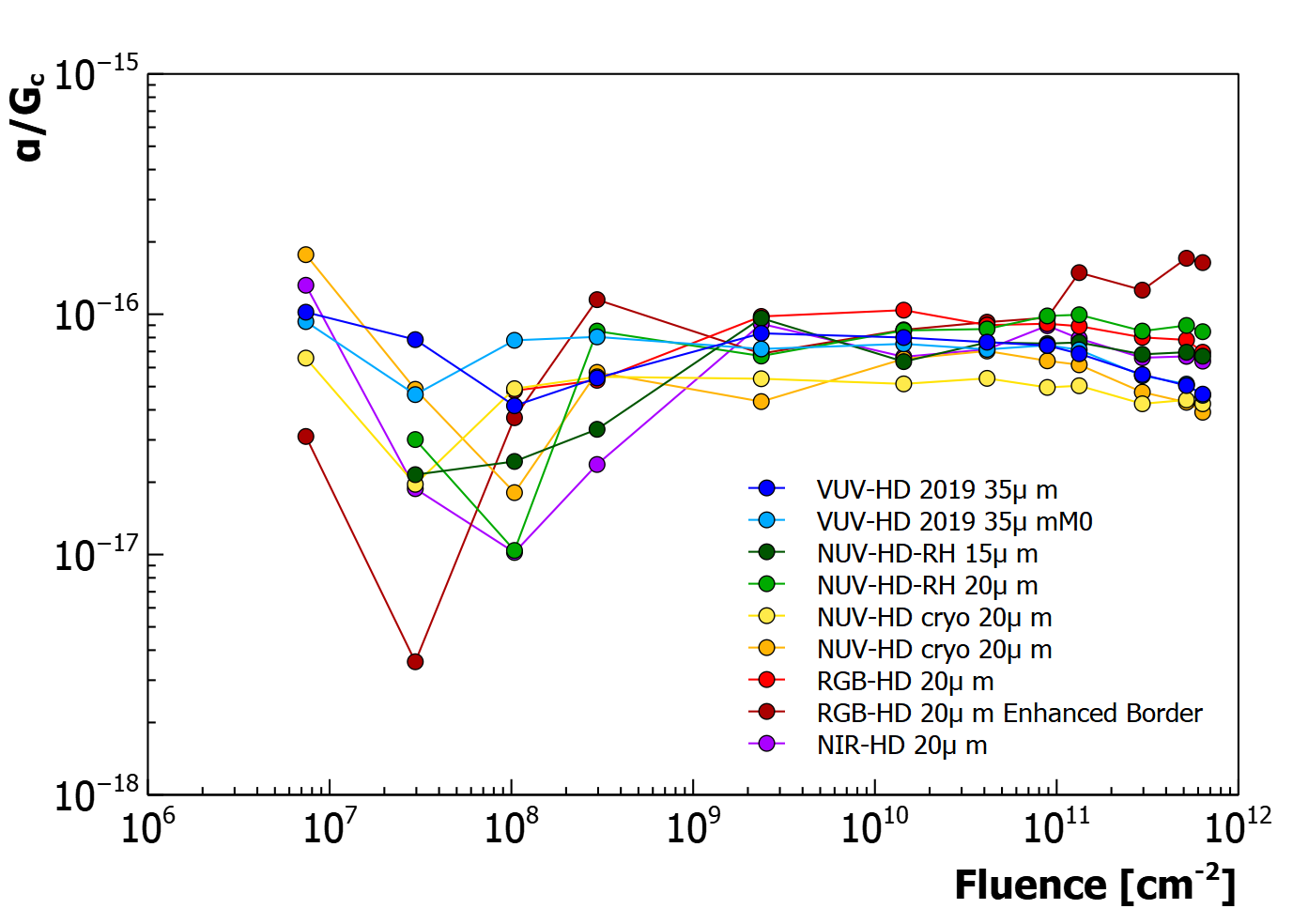}
    \caption{Damage parameter of the tested technologies as a function of the fluence at 3V excess bias.}
    \label{fig:alfa}
\end{figure}

The increase of the current $\Delta I = I_{\phi}-I_{\phi_0}$ in dark conditions was used to estimate the \textit{damage parameter} $\alpha$ \cite{Moll}. To do this, the current increase was normalized to the \textit{current gain} \cite{altamura2021characterization}, the product of the micro-cell gain and the excess charge factor (ECF), which represents the ratio between the average charge produced by one primary event (including all correlated secondary pulses generated by cross-talk and afterpulsing) and the charge of just one primary event \cite{PIEMONTE-gola}. This parameter gives an idea of the effective damage inside the SiPM and provide a first comparison among different technologies.  
In Fig.\ref{fig:alfa}, the damage parameter $\alpha$ is plotted as a function of the fluence at 3V excess bias, showing a fairly uniform trend of all the tested technologies, apart form the RGB-HD SiPM with Enhanced border 20 $\mu m$ cell, which appears more affected by the radiation than the other SiPMs, showing an increase of $\alpha$ starting from $10^{11}$ $n_{eq}/cm^2$. 

\begin{figure}[tb]
    \centering
    \includegraphics[width=0.45\textwidth]{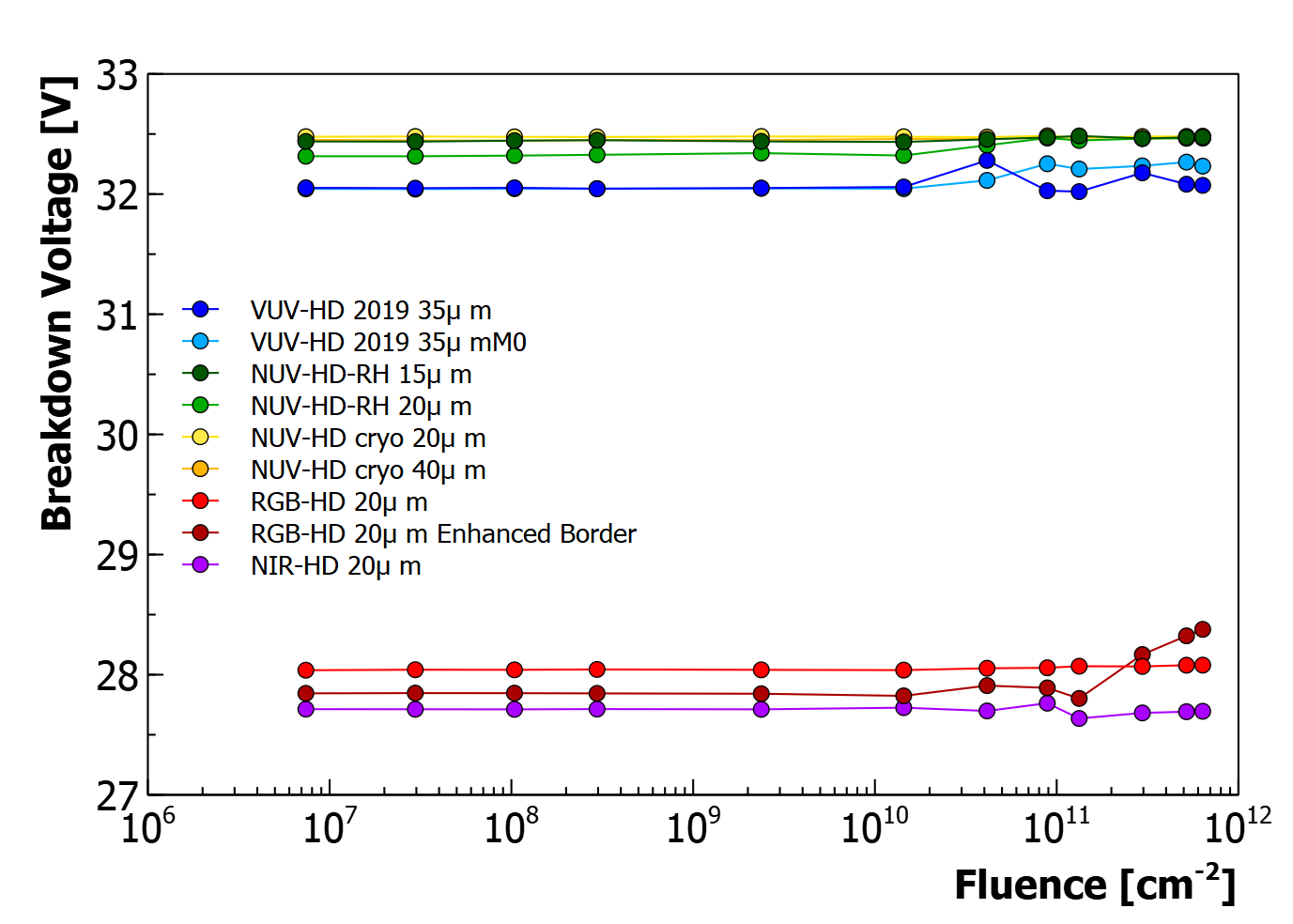}
    \caption{Breakdown voltage of the tested technologies as a function of the fluence.}
    \label{fig:Vbd}
\end{figure}

From the reverse current measurement in light conditions, the breakdown voltage was estimated from the second logarithmic derivative of the reverse current, as visible in Fig.\ref{fig:Vbd}. The breakdown voltage appears to remain constant for almost all the technologies apart from the RGB-HD SiPM with Enhanced border 20 $\mu m$ cell, where it starts to increase at $10^{11}$ $n_{eq}/cm^2$. This could mean that, at least in this latter technology, the implant doping concentration of the single SPADs were damaged by the radiation, while this does not hold true for the other technologies.

\subsection{PDE}

\begin{figure}[tb]
    \centering
    \includegraphics[width=0.45\textwidth]{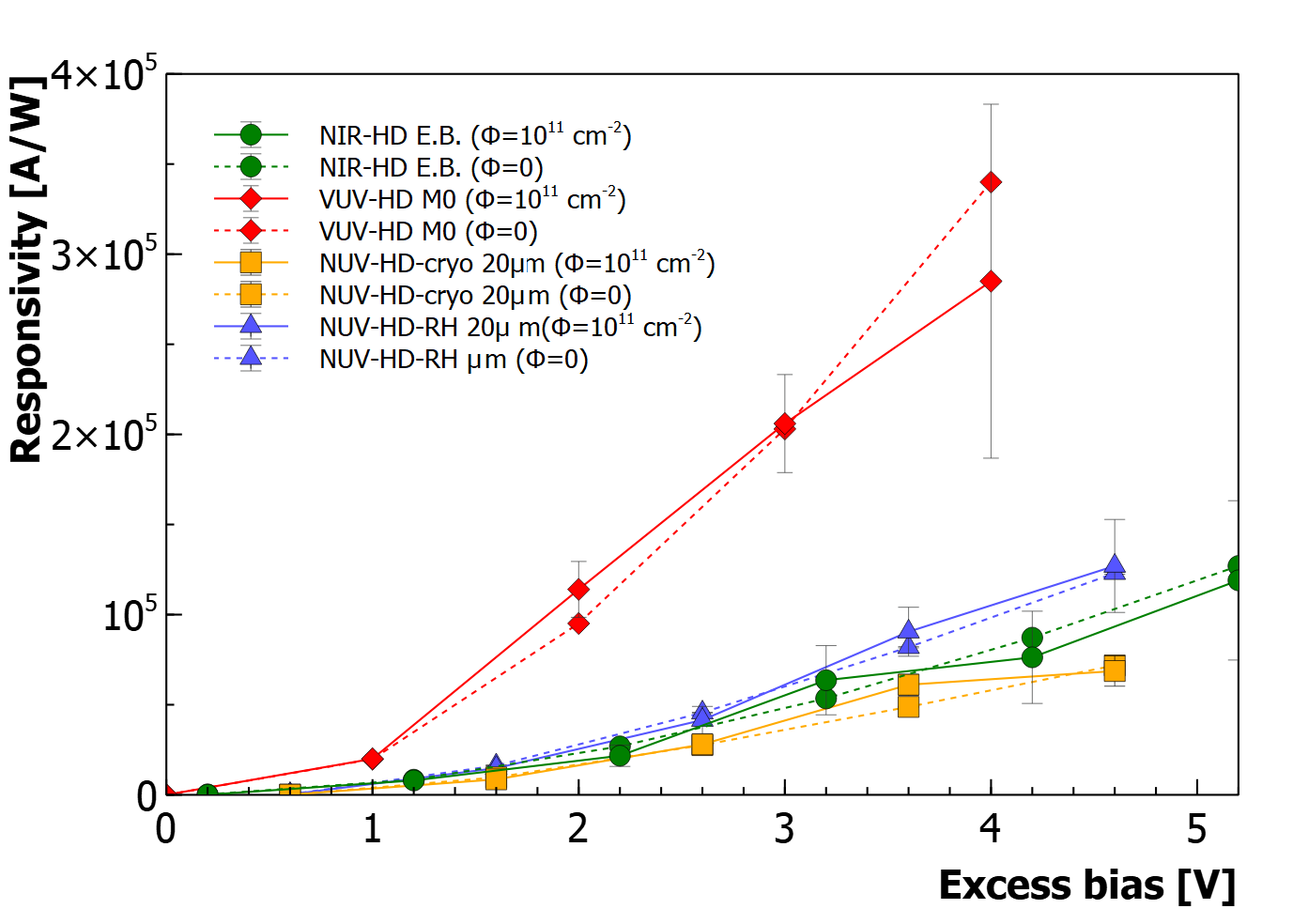}
    \caption{Responsivity as a function of the excess bias for RGB-HD, NUV-HD-RH, VUV-HD 2019 and NUV-HD-cryo technologies, irradiated at $10^{11}$ $n_{eq}/cm^2$ (straight line) and not-irradiated (dashed line).}
    \label{fig:Responsivity}
\end{figure}

Photon Detection Efficiency (PDE) of the SiPMs is crucial to investigate their ability in photo-detection. In highly-irradiated SiPMs, a direct measurement of the PDE is not possible due to the increase of the dark current which becomes comparable to the light current when the SiPMs are illuminated, thus the measurement is no longer accurate.  
For this reason, the results on the PDE were derived by a \textit{Responsivity} measurement. That was performed on the samples irradiated at $10^{11}$ $n_{eq}/cm^2$, which were illuminated by a 420 nm LED. The results are visible in Fig.\ref{fig:Responsivity}, where the responsivity is plotted as a function of the excess bias, for irradiated (straight line) and not-irradiated (dashed line) samples. One sample for each technology was considered, and no significant changes were observed in any of them. This means that the PDE remains constant with fluence, at least up to $10^{11}$ $n_{eq}/cm^2$ in all the technologies under test.

\subsection{DCR}

Dark Count Rate is one of the most important parameters in a SiPM. A direct measurement of the DCR in highly-irradiated devices can be very difficult, because of the high count rate which prevents the use of the common DCR characterization methods based on a waveform analysis. For this reason, we chose to extract the DCR from the dark current and the so-called \textit{Current Gain} $G_c$, which was previously introduced. The DCR can be estimated as:
\begin{equation}
    DCR=\frac{1}{q}\frac{I_{dark}}{G_c}
    \label{eq:Gc}
\end{equation}

\begin{figure}[tb]
    \centering
    \includegraphics[width=0.45\textwidth]{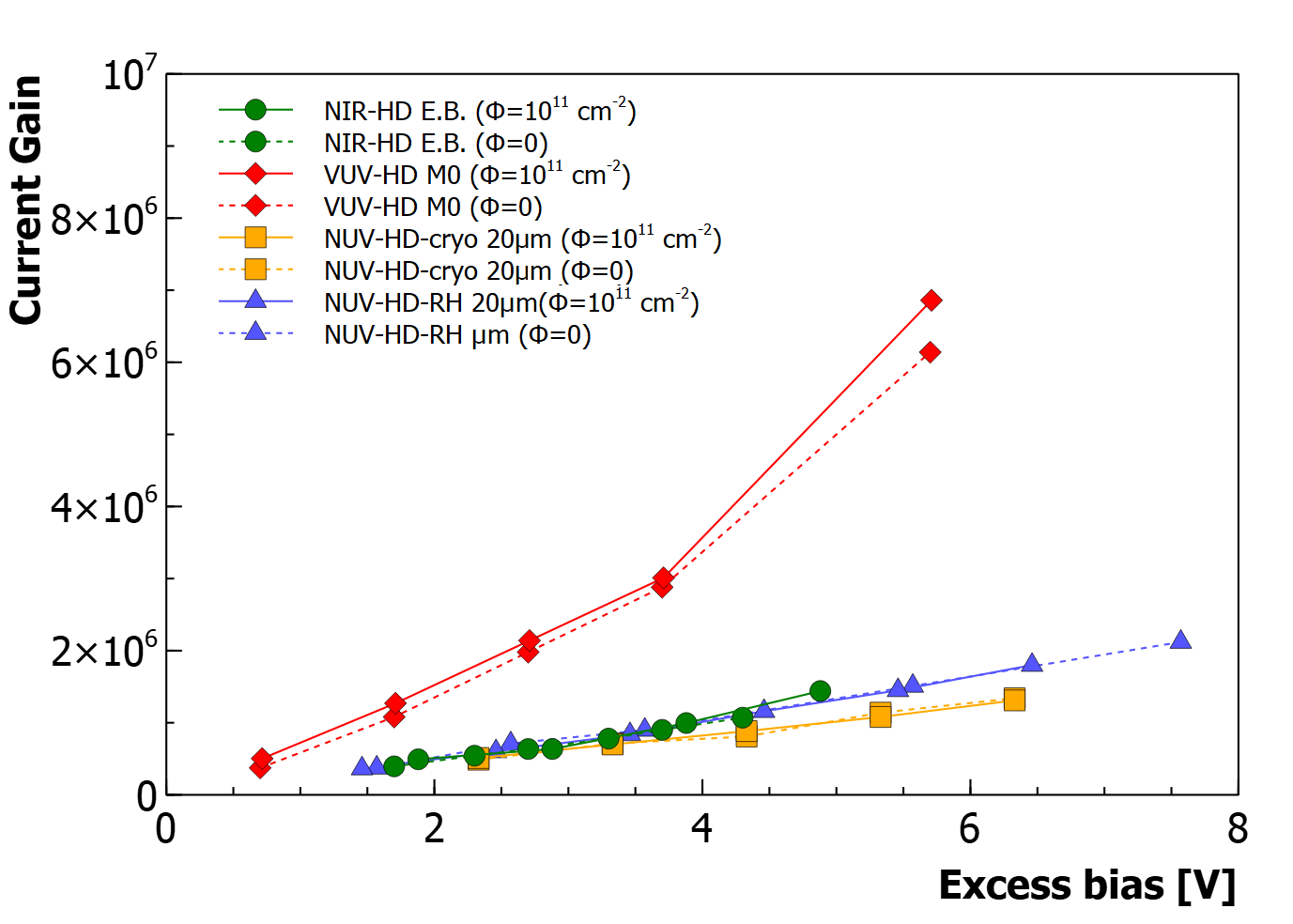}
    \caption{Gain current as a function of the excess bias mneasured at -40$^{\circ}$C for the RGB-HD, NUV-HD-RH, VUV-HD 2019 and NUV-HD-cryo technologies, irradiated at $10^{11}$ $n_{eq}/cm^2$.}
    \label{fig:Gc}
\end{figure}

In irradiated SiPMs, Eq.\ref{eq:Gc} holds under the assumption that the current gain measured before irradiation remains constant with the fluence\cite{altamura2021characterization}. To verify this assumption, we performed a direct $G_c$ measurement on the samples irradiated at $10^{11}$ $n_{eq}/cm^2$ at -40$^{\circ}$C, due to the high noise rate which did not make a direct measurement at lower temperature possible. The results are visible in Fig.\ref{fig:Gc}, where the current gain is plotted as a function of the excess bias.  
The plot does not shown any relevant changes, suggesting that the current gain is not affected by the radiation, at least up to $10^{11}$ $n_{eq}/cm^2$. 

\begin{figure}[tb]
    \centering
    \includegraphics[width=0.45\textwidth]{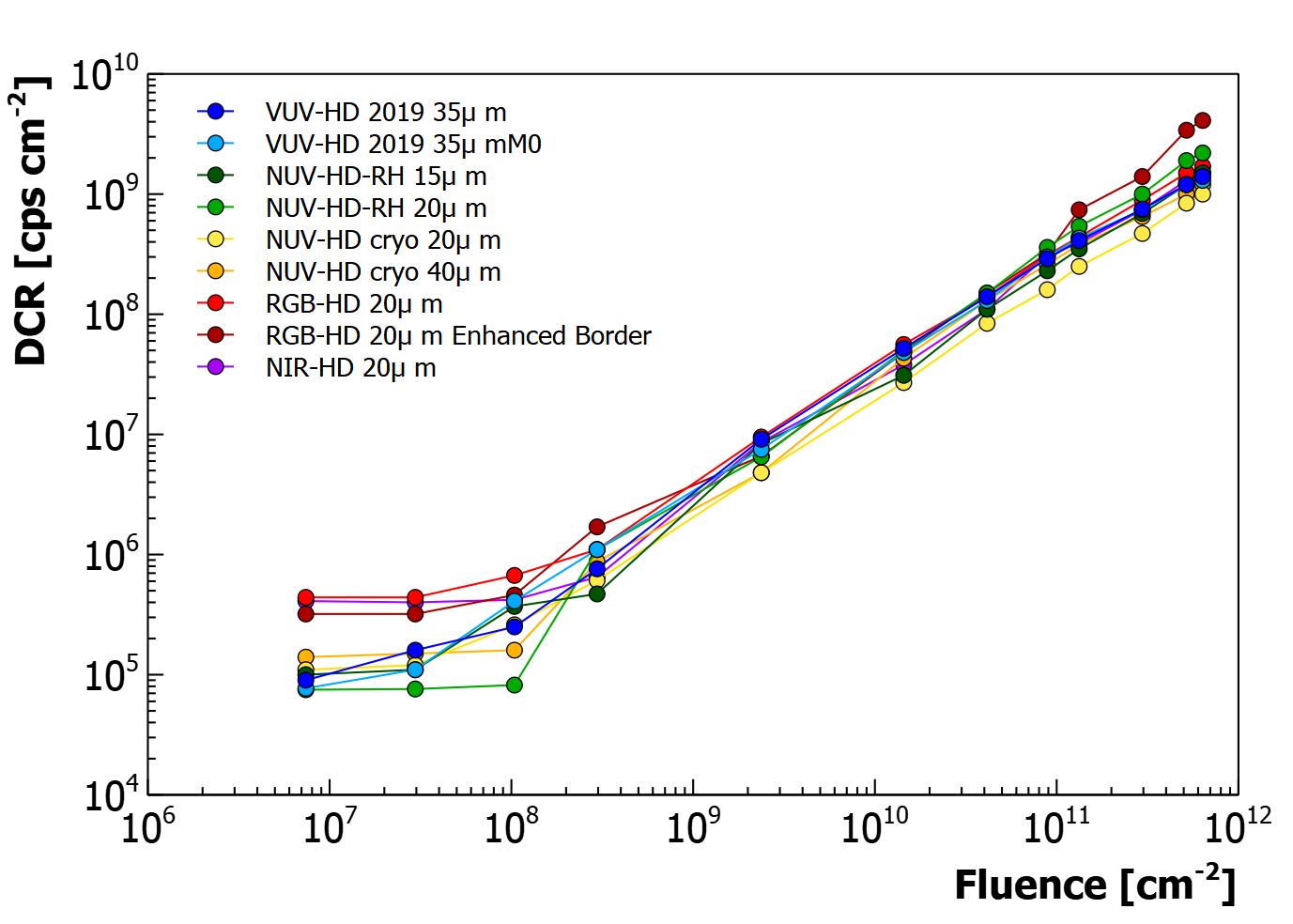}
    \caption{Dark Count Rate of the tested SiPMs as a function of the irradiation fluence.}
    \label{fig:DCR}
\end{figure}

After verifying the current gain does not change with fluence, the DCR was extracted at all the fluences for all the technologies under test, as shown in Fig.\ref{fig:DCR}, where the curves of the several devices look more similar as the fluence increases, not showing any significant trend. The RGB-HD SiPM with Enhanced border 20 $\mu m$ cell appears more damaged than the others, starting from $10^{11}$ $n_{eq}/cm^2$, as already observed in the case of the damage parameter $\alpha$ and the breakdown voltage. 

Although not obvious from Fig.\ref{fig:DCR}, some SiPMs, mainly the larger-sized ones, exhibit some saturation effects at high fluences. SiPM saturation (due to the high cell occupancy) means that the DCR values are comparable to the maximum possible DCR of a SiPM with all cells constantly triggering\cite{altamura2021characterization} and it appears as a anomalous deflection. 

\begin{figure}[tb]
    \centering
    \includegraphics[width=0.45\textwidth]{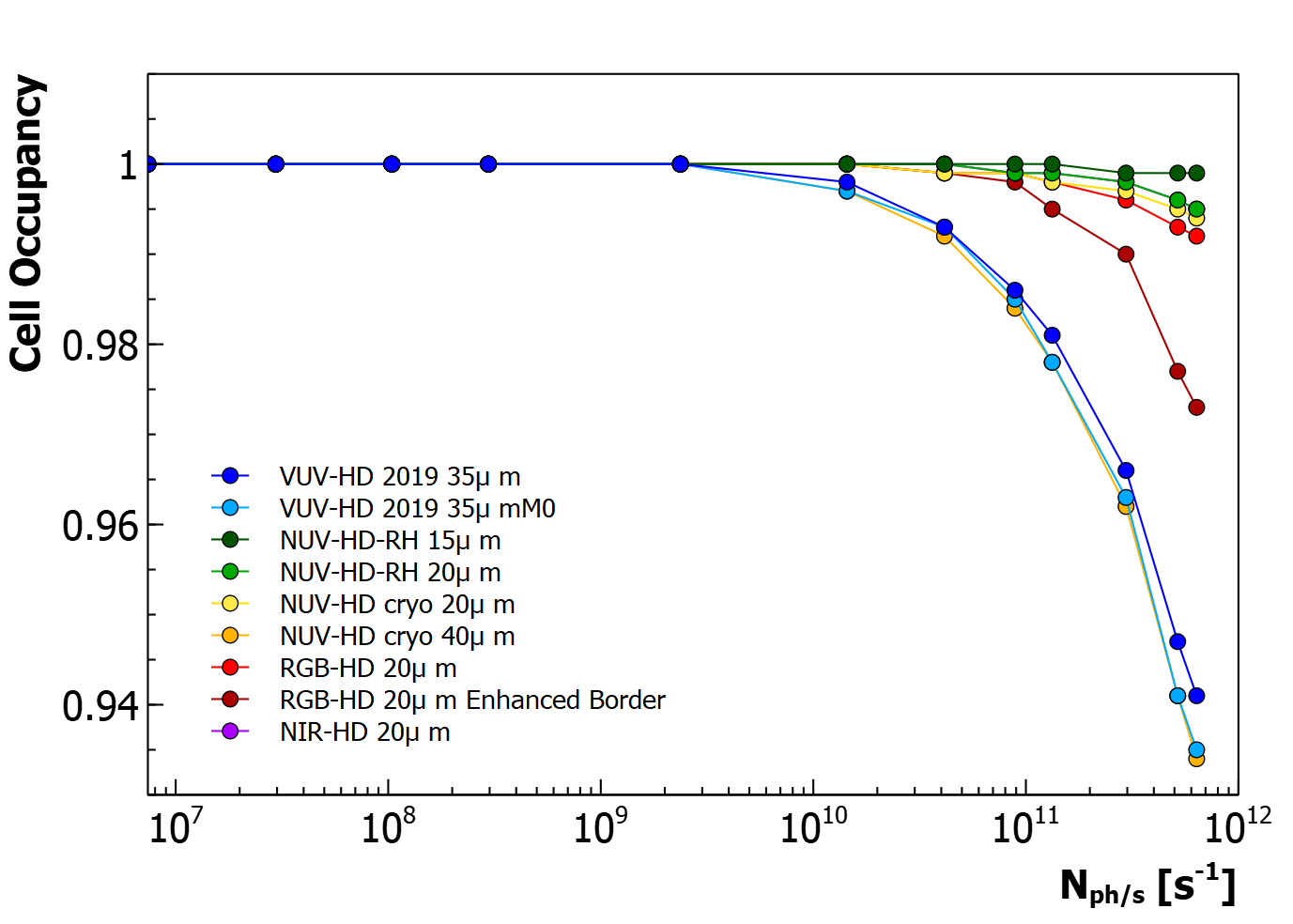}
    \caption{Saturation of SiPM cells as a function of the fluence at 3V excess bias, with visible effects on the devices with larger cell pitches.}
    \label{fig:saturation}
\end{figure}

In Fig.\ref{fig:saturation} a plot of the SiPM saturation (or cell occupancy) as a function of the fluence at 3V excess bias is visible, where we observe a more clear deflection of the curve for large-sized cell SiPMs, whereas the curve of small-sized cell SiPMs, remains almost constant.

\subsection{SNR and energy resolution}
The results shown above are typical functional parameters of SiPMs. In order to compare the technologies under test and select the most promising ones for space applications, other parameters should be considered to extract figure of merits more related to experiments operating in space. 


An example of figure in merit useful to compare the SiPMs performance before and after irradiation is the detector Signal-to-Noise (SNR) ratio which, for pulse counting detectors, is defined as:

\begin{equation}
    SNR = \frac{N_t^{tot}-N_t^{noise}}{\sqrt{ENF\cdot N_t^{tot}}}
\end{equation}

where ENF is the Excess Noise Factor, i.e. a measure of the "inaccuracy" of the sensor and:
\small
\begin{align}
    N_t^{tot} &= N_{cell} \frac{t_{int}}{\tau} \Big(1-e^{-\frac{1+CT}{N_{cell}} (N_{ph/s}\cdot t_{int} \cdot PDE + DCR \cdot t_{int})\frac{\tau}{t_{int}}}\Big) \\ 
    N_t^{noise} &= N_{cell} \frac{t_{int}}{\tau} \Big(1-e^{-\frac{1+CT}{N_{cell}} (DCR \cdot t_{int})\frac{\tau}{t_{int}}}\Big) 
\end{align}
\normalsize

where $N_t^{tot}$ is the number of SPADs triggered by the total light and noise events in a certain integration time $t_{int}$, $N_t^{noise}$ is the number of SPADs triggered by noise events in the same integration time, $N_{ph/s}$ is the number of incident photons per second and $\tau$ is the recharge time of the cells. 

\begin{figure}[tb]
    \centering
    \includegraphics[width=0.45\textwidth]{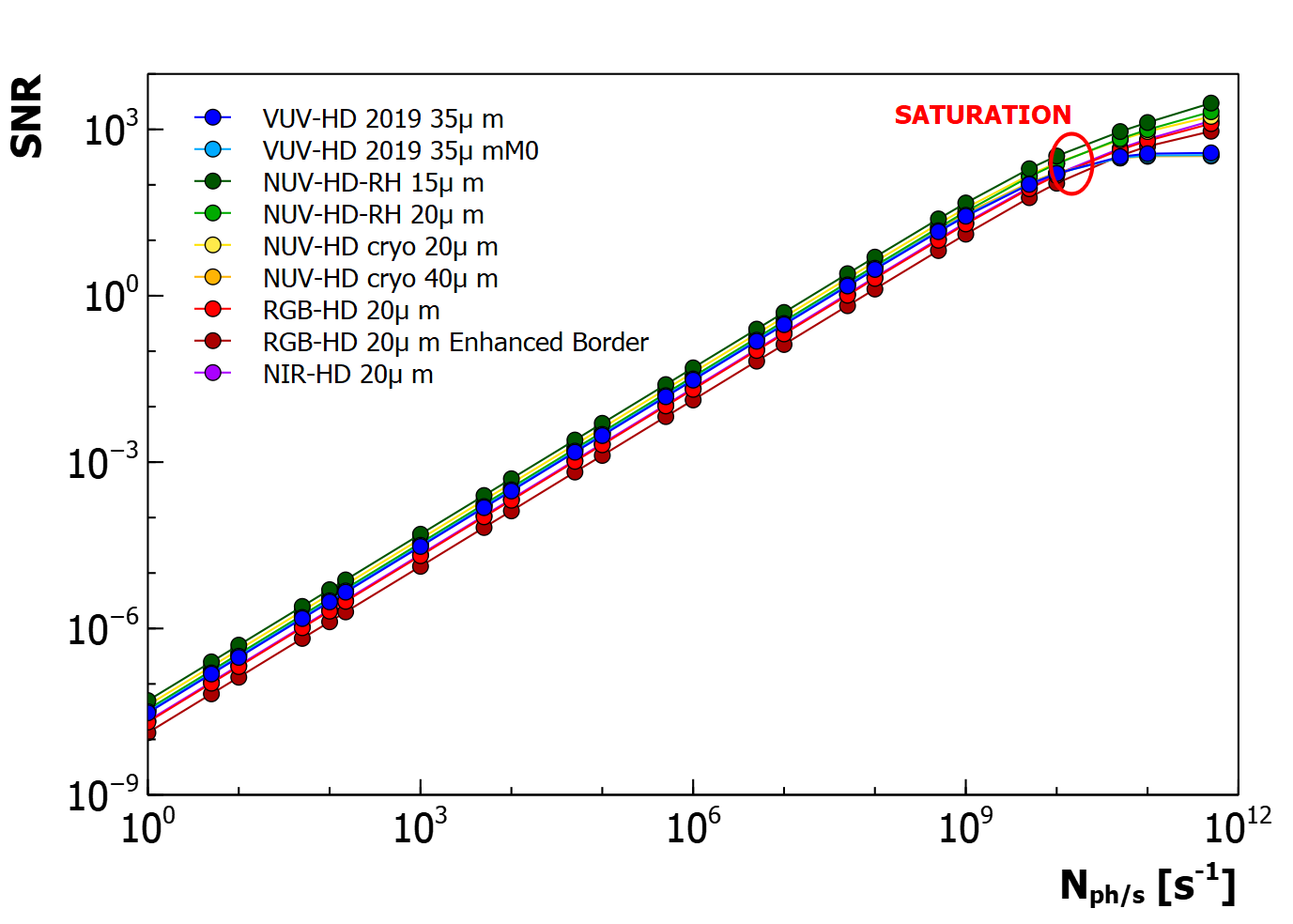}
    \caption{Signal-to-noise ratio as a function of the fluence at fixed excess bias (3V), integration time ($6.8\times10^{-8}$ s) and fluence ($6.4\times 10^{11}$ $n_{eq}/cm^2$).}
    \label{fig:SNRvsph}
\end{figure}

Fig.\ref{fig:SNRvsph} shows the trend of the SNR as a function of the number of incident photons, at fixed integration time ($2\times 10^{-6}$) and fluence ($6.4\times 10 ^{11}$ $n_{eq}/cm^2$). Here we observe a better performance of the large-sized SiPMs due to their high PDE up to $10^{10}$ ph/s, where they start to suffer a saturation process, whereas the small-sized SiPMs appear less affected by the saturation.  

In order to extract quantitative values, thus compare the SiPM technologies, we needed to target a specific case. For example, the number of incident 450nm photons per second was fixed at $10^{10}$ $ph/s$ to get an high SNR without considering saturation effects, whereas the integration time was set to $2\times 10^{-6} s$ as in the ALPIDE Monolithic Active Pixel Silicon sensor\cite{AglieriRinella:2239752}, developed at CERN for the ALICE ITS Upgrade and also considered for future satellite particle trackers.

\begin{figure}[tb]
    \centering
    \includegraphics[width=0.45\textwidth]{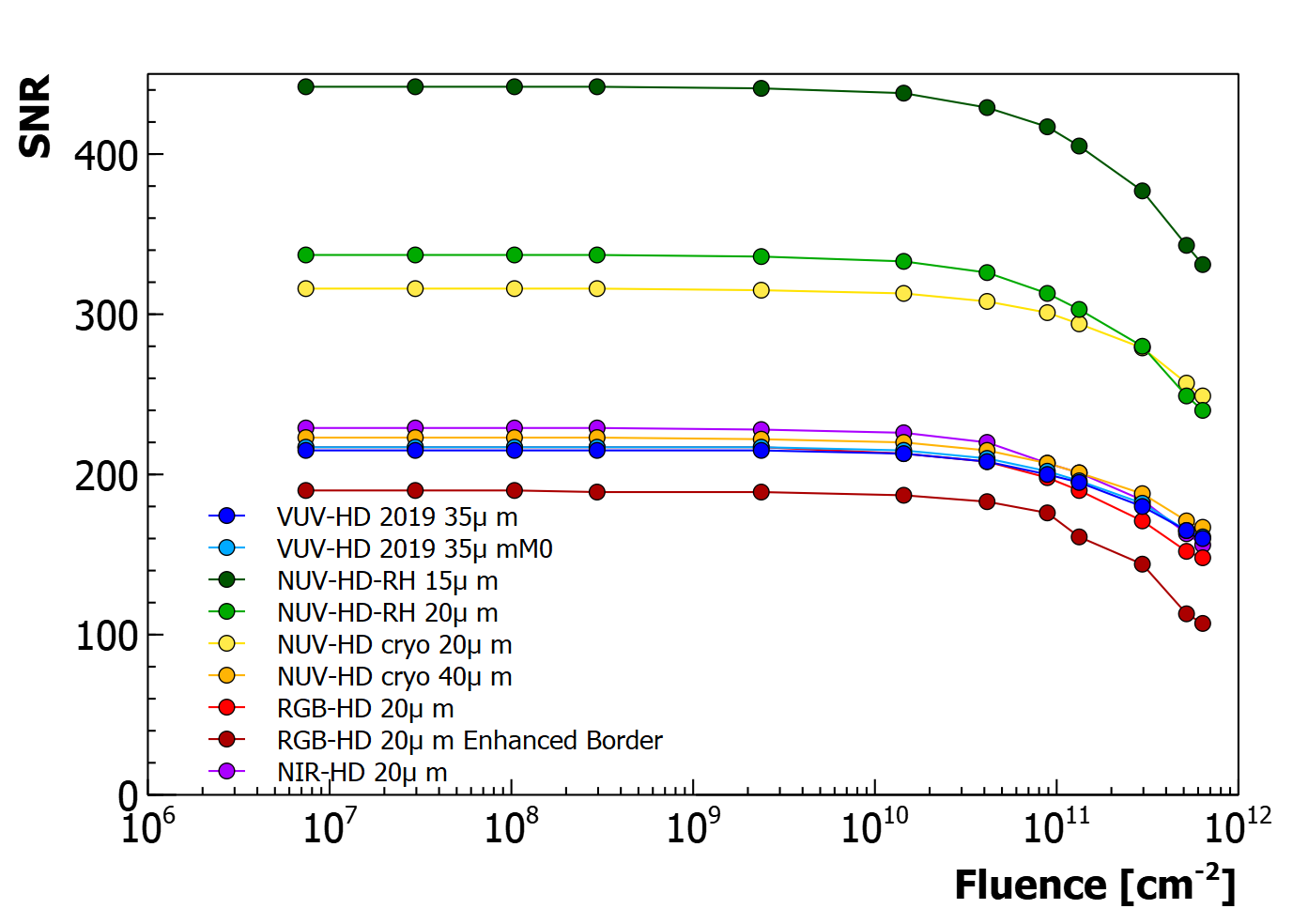}
    \caption{Signal-to-noise ratio as a function of the fluence at fixed excess bias (3V), integration time ($6.8\times10^{-8}$ s) and number of photons ($2\times 10^{10}$).}
    \label{fig:SNRvsF}
\end{figure}

In Fig.\ref{fig:SNRvsF} the SNR curves are plotted as a function of the fluence at fixed number of incident photons per second ($10^{10}$ $ph/s$) and integration time window ($2\times 10^{-6} s$). Here we can see a general worsening of the performance of the sensors at around $10^{11}$ $n_{eq}/cm^2$. 
Among the SiPMs with 20 $\mu$m cell pitch, we observe a low SNR (around 200) for the NIR-HD, NUV-HD cryo and RGB-HD technologies, whereas the NUV-HD-RH and NUV-HD-cryo with 20 $\mu$m cell pitch technologies show a better performance. In particular, the most significant result comes from the NUV-HD-RH SiPM with 15 $\mu$m cell which show a high SNR at at 440. 
Among the large-sized SiPMs, which are supposed to have an higher SNR due to their high FF and PDE but also an earlier saturation, the NUV-HD-cryo technology with 40 $\mu$m cell shows the highest SNR. 

In X or $\gamma$-ray spectroscopy, the energy resolution should be taken into account:

\small
\begin{equation}
    \sigma_E \propto \sqrt{\frac{ENF^*}{N_{ph/s}\cdot t_{int} \cdot PDE}+\frac{ENF^* \cdot DCR  \cdot t_{int}}{(N_{ph/s} \cdot t_{int} \cdot PDE)^2}}
    \label{eq:ER}
\end{equation}
\normalsize
 
where ENF$^*$ is the Excess Noise Factor, in this case estimated as product of several contributions, as in \cite{ENF}.

\begin{figure}[tb]
    \centering
    \includegraphics[width=0.45\textwidth]{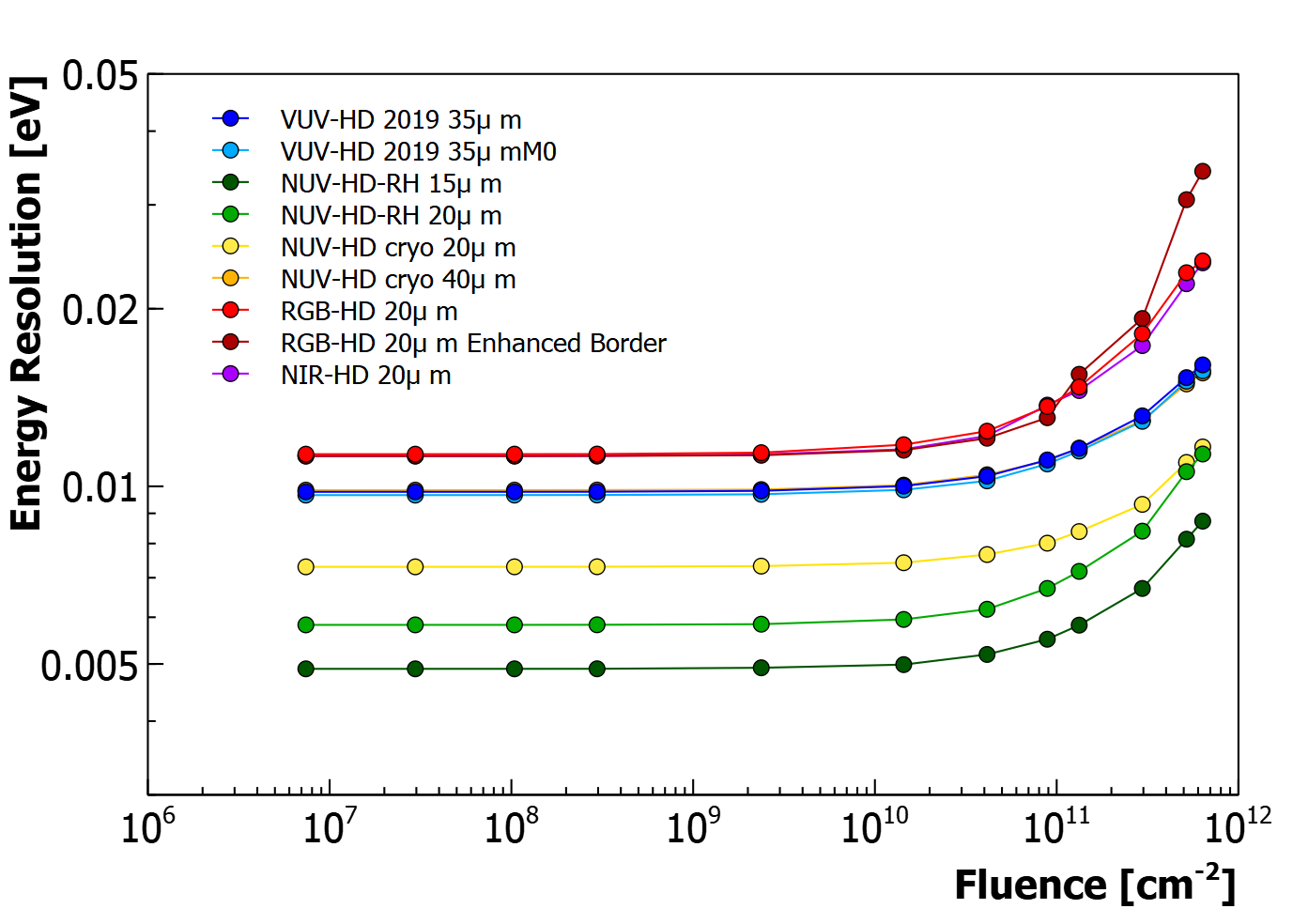}
    \caption{Energy resolution as a function of the fluence at fixed excess bias (3V), integration time ($6.8\times10^{-8}$ s) and number of photons ($2\times 10^{10}$).}
    \label{fig:ER}
\end{figure}

The results of the energy resolution estimation are visible in Fig.\ref{fig:ER}, where a clear distinction between the SiPMs with small (15-20 $\mu$m) and large (35-40 $\mu$m) cell pitches can be noted. In particular, the large-sized ones have an higher $\sigma_E$, at around 1$\%$. A trend with the cell size can be observed, apart form the RGB-HD SiPM with Enhanced border 20 $\mu m$ cell which has an higher $\sigma_E$, nearly 1$\%$, at the same level of the 35 $\mu$m cell SiPMs. Furthermore, it starts to have a dramatic increase at $10^{11}$ $n_{eq}/cm^2$ until reaching a 2$\%$ value.

Overall, NUV-HD-RH and NUV-HD-cryo technologies show the best performance under the conditions described above after irradiation, with an higher SNR and a lower energy resolution. It has to be pointed out that this result is merely related to the application that we took into account, in this particular case the X and $\gamma$-ray spectroscopy. 

\subsection{Annealing}
\begin{figure}[tb]
    \centering
    \includegraphics[width=0.45\textwidth]{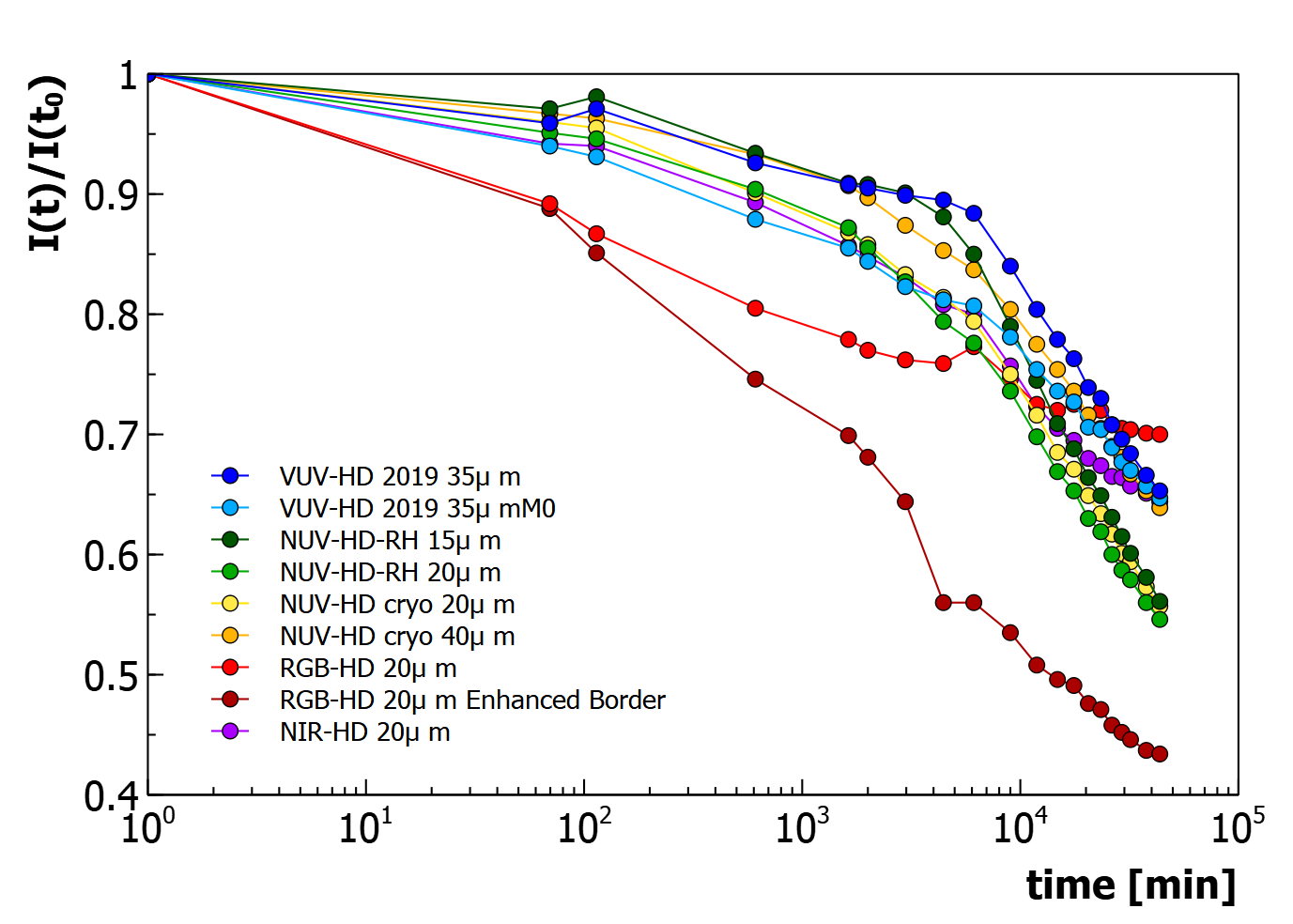}
    \caption{$I(t)/I(t_0)$ ratio as a function of the annealing time after irradiation at room temperature ($20\div 25^{\circ}C$).}
    \label{fig:Annealing}
\end{figure}

A 30-days annealing test on all the irradiated SiPMs was performed at room temperature ($20\div25^{\circ}C$), with reverse current measurements twice per day. The normalized dark currents are plotted in Fig.\ref{fig:Annealing} at 3V of excess bias. 
Here, the RGB-HD SiPMs appear to have the fastest recovery. Indeed, the RGB-HD SiPM with Enhanced border $20\mu m$ cell reduces its current of almost a factor 0.4 after the whole annealing time, while the RGB-HD SiPM with standard 20 $\mu m$ cell, after a fast initial recovery starts to slow after 100 minutes. The other technologies show an initial constant decrease until $10^3$ minutes, where they present a knee and then a faster recovery. 
As visible in the plot, after the whole 30-days annealing time, the RGB-HD Enhanced Border, NUV-HD-RH and NUV-HD-cryo 20$\mu$m technologies are the ones recovering the most, while the VUV-HD and NUV-HD-cryo 40$\mu$m appear to recover the less. 

\section{Conclusions}
An irradiation test was performed at the Trento Proton Therapy Center in February 2021. Several FBK technologies with different cell piches and shapes were irradiated with 74 MeV protons at gradual fluence steps, up to $6.4\times10^{11}$ $n_{eq}/cm^2$. 

After a reverse current measurement, the damage factor $\alpha$ was extracted. This did not show any clear trend for all the technologies apart from the RGB-HD SiPM with Enhanced border 20 $\mu m$ cell, which appears to be more damaged starting from $10^{11}$ $n_{eq}/cm^2$. The same behaviour can be observed for the breakdown voltage, where the RGB-HD SiPM with Enhanced border 20 $\mu m$ cell is the only technology to suffer from the irradiation.
The DCR in all the different technologies starts at different levels but slowly catch each other up at increasing fluences, until reaching similar values. The RGB-HD SiPM with Enhanced border 20 $\mu m$ cell shows a deflection at $10^{11}$ $n_{eq}/cm^2$, which is a result consistent with what observed for the damage parameter and the breakdown voltage. At high fluences some of the SiPMs under test present a saturation effect, which is particularly prominent for large-sized SiPMs and, nonetheless, it is a parameter to take into account during the whole characterization process. 

An effective comparison among the technologies can be achieved through the estimation of the energy resolution and the signal-to-noise ratio, which provide information about the response of the sensor when exposed to a photon flux. These two parameters were estimated taking into account the saturation of the cell and the total excess noise factor from different noise sources. The SNR is used in this work as a figure of merit. It shows a trend with the cell size at fixed fluence. In particular, at $6.4\times10^{11}$ $n_{eq}/cm^2$ the larged-sized SiPMs exhibit an higher SNR with a clear saturation starting at at around $10^{10}$ ph/s. At fixed number of incident photons per second, the NUV-HD-RH 20 $\mu$m and NUV-HD-cryo technologies show the highest SNR values, in case of small cells.  

After irradiation, a 30-days annealing at room temperature was performed on the SiPMs. The RGB-HD SiPM with Enhanced border 20 $\mu m$ cell showed a faster recovery, whereas the other technologies exhibited a very similar behaviour with a constant decrease followed by a knee and a faster decrease, up to around 50$\%$ recovery factor.

This work had as a main purpose a very preliminary characterization of the FBK SiPMs for space applications. To do that, typical fluences experienced by satellites in space were considered. All the SiPMs survived to the irradiation tests and only the RGB-HD SiPM with Enhanced border 20 $\mu m$ cell was observed to deviate significantly from the behaviour of the other technologies at fluences above $10^{11}$ $n_{eq}/cm^2$. 
Overall, due to the higher SNR and the fast recovery during the annealing time, the NUV-HD-RH and the NUV-HD-cryo 20 $\mu$m technologies showed the best performance. For the NUV-HD-RH, two cell sizes were tested and the 15 $\mu$m SiPM displayed a better behaviour than the 20 $\mu$m SiPM.





\bibliographystyle{model1-num-names}
\bibliography{sample.bib}







\end{document}